\documentclass[iop,tighten]{emulateapj}

\usepackage{apjfonts}

\newcommand{\msun}{M_\odot}
\newcommand{\src}{G1.9+0.3}

\newcommand{\scand}{$^{44}$Sc}
\newcommand{\ti}{$^{44}$Ti}

\catcode`\@=11
\newcommand{\gapprox}{\mathrel{\mathpalette\@versim>}}
\newcommand{\lapprox}{\mathrel{\mathpalette\@versim<}}
\newcommand{\propapprox}{\mathrel{\mathpalette\@versim\propto}}
\newcommand{\@versim}[2]
  {\lower3.1truept\vbox{\baselineskip0pt\lineskip0.5truept
\ialign{$\m@th#1\hfil##\hfil$\crcr#2\crcr\sim\crcr}}}
\catcode`\@=12
\journalinfo{ApJ Letters, in press}
\submitted{}
\shorttitle{RADIOACTIVE SCANDIUM IN YOUNGEST GALACTIC SNR G1.9+0.3}

\begin{document}

\title{Radioactive Scandium in the  
Youngest Galactic Supernova Remnant G1.9+0.3}

\author{Kazimierz J. Borkowski,\altaffilmark{1}
Stephen P. Reynolds,\altaffilmark{1}
David A. Green,\altaffilmark{2}
Una Hwang,\altaffilmark{3}
Robert Petre, \altaffilmark{3}
Kalyani Krishnamurthy, \altaffilmark{4}
\&\ Rebecca Willett \altaffilmark{4}
}

\altaffiltext{1}{Department of Physics, North Carolina State University, 
Raleigh, NC 27695-8202; kborkow@unity.ncsu.edu} 
\altaffiltext{2} {Cavendish Laboratory; 19 J.J. Thomson Ave., 
Cambridge CB3 0HE, UK}
\altaffiltext{3}{NASA/GSFC, Code 660, Greenbelt, MD 20771}
\altaffiltext{4}{Department of Electrical and Computer Engineering, 
Duke University, Durham, NC 27708}

\begin{abstract}

We report the discovery of thermal X-ray emission from the youngest
Galactic supernova remnant G1.9+0.3, from a 237-ks {\sl Chandra}
observation.  We detect strong K$\alpha$ lines of Si, S, Ar, Ca, and
Fe.  In addition, we detect a 4.1 keV line with 99.971\% confidence
which we attribute to $^{44}$Sc, produced by electron capture from
$^{44}$Ti.  Combining the data with our earlier {\sl Chandra}
observation allows us to detect the line in two regions independently.
For a remnant age of 100 yr, our measured total line strength
indicates synthesis of $(1 - 7) \times 10^{-5} \msun$ of \ti, in the
range predicted for both Type Ia and core-collapse supernovae,
but somewhat smaller than the $2 \times 10^{-4} \msun $ reported for
Cas A.  The line spectrum indicates supersolar abundances.  The Fe
emission has a width of about 28,000 km s$^{-1}$, consistent with an
age of $\sim 100$ yr and with the inferred mean shock velocity of
14,000 km s$^{-1}$ deduced assuming a distance of 8.5 kpc.  Most
thermal emission comes from regions of lower X-ray but higher radio
surface brightness.  Deeper observations should allow more detailed
spatial mapping of \scand, with significant implications for models of
nucleosynthesis in Type Ia supernovae.

\end{abstract}


\section{Introduction}
\label{intro}

The supernova remnant (SNR) G1.9+0.3 has an age of order 100 years
\citep{reynolds08b,green08}.  Its integrated X-ray spectrum is
well described by a model of synchrotron emission from a power-law
distribution with an exponential cutoff (XSPEC model {\tt srcut}),
with rolloff frequency $h\nu_{\rm roll} = 2.2$ keV and a very high absorbing
column density $N_H \cong 5 \times 10^{22}$ cm$^{-2}$ \citep[][Paper
II]{reynolds09}.  The high column suggests a distance of order that to
the Galactic Center; we adopt a nominal distance of 8.5 kpc, at which
the mean expansion rate is $v_{\rm s} \sim 14,000$ km s$^{-1}$ (Paper
I).  No thermal emission is apparent in the integrated spectrum based on 
short (50 ks in duration) {\it Chandra} observations in early 2007.
Without the detection of thermal emission, basic, crucial information such as 
the supernova (SN) type, distance, and elemental abundances 
cannot be obtained.  Its detection was one of the
goals of a much longer {\sl Chandra} observation.

\section{Observations}
\label{obssec}

We re-observed \src\ for 237 ks with {\sl Chandra} in four
observations between July 13 and 26, 2009, using the ACIS-S CCD camera
(S3 chip).  We checked aspect correction and created new level--1
event files appropriate for VFAINT mode. No flares occurred during the
observation.  CTI correction was applied and calibration was performed
using CALDB version 3.4.0.  Finally, the datasets were
merged and weighted response files created. We extracted spectra using
the {\tt specextract} script.
We obtained about 40,000 source counts.

\begin{figure}
\epsscale{1.1}
\vspace{0.1truein}
%\plotone{f1.eps}
%\plotone{3colorb.eps}
%\plotone{g1p9threecolor4b.jpg}
\plotone{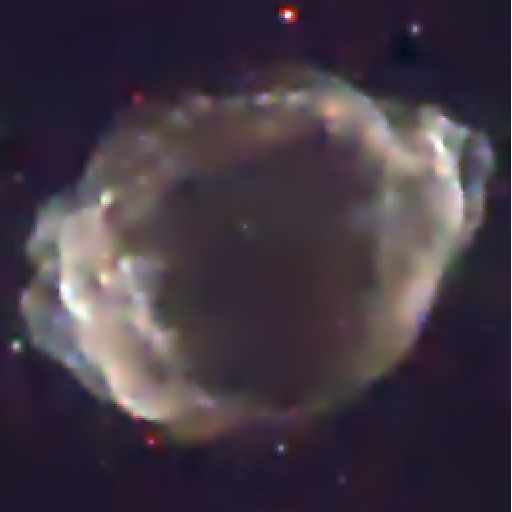}
\caption{
{\sl Chandra} image of \src. 
Red,
1 -- 3 keV; green, 3 -- 4.5 keV; blue, 4.5 -- 7.5 keV.
Image size $127'' \times
121''$.
\label{2xim}}
\end{figure}

\begin{figure}
\epsscale{1.1}
%\plotone{radio.jpg}
%\plotone{g1p9tworegions.jpg}
%\plotone{radio.eps}
%\plotone{g1p9tworegions.eps}
%\plotone{f2.eps}
\plotone{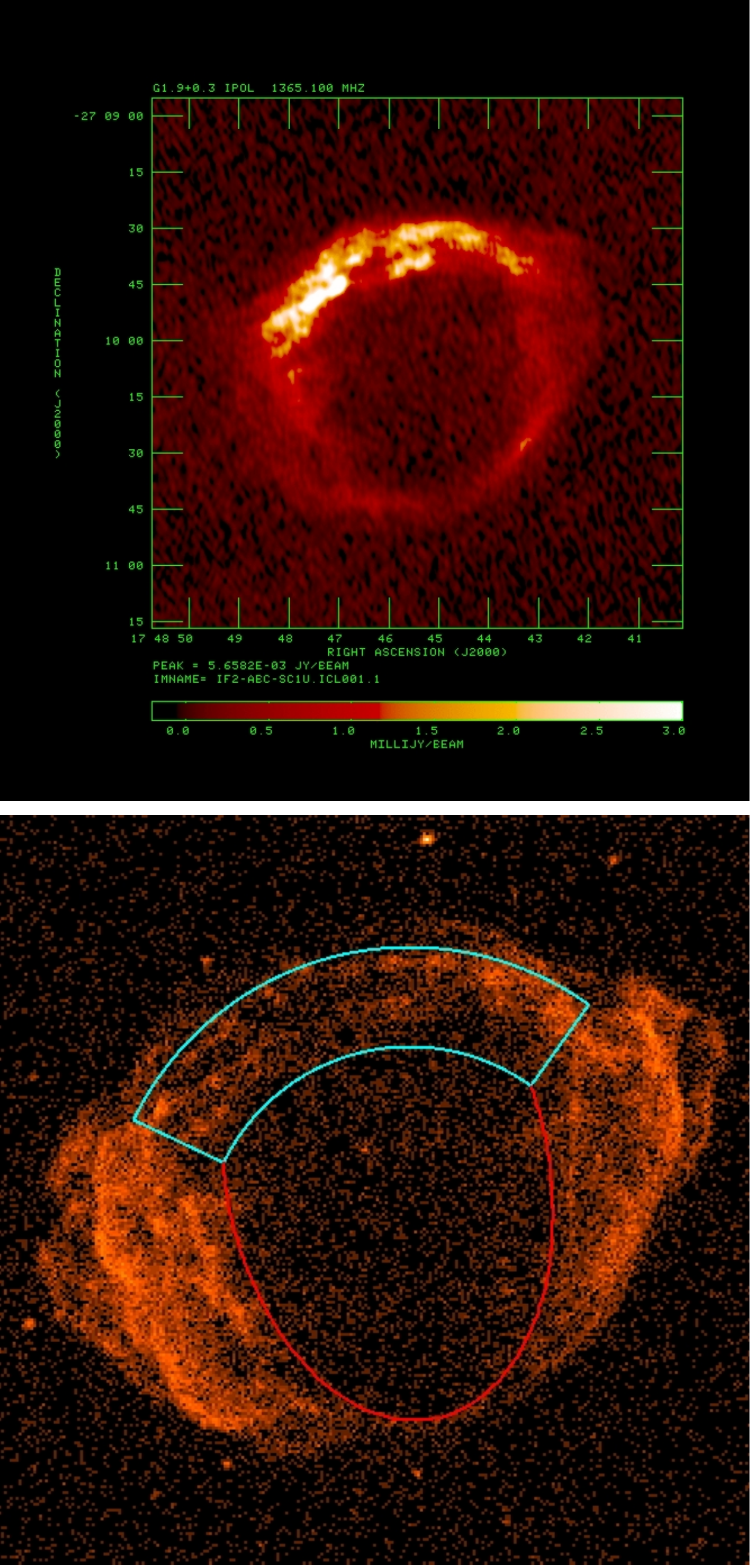}
\caption{Top: 1.4 GHz radio image (Green et al.~2010, in preparation).
Resolution $2.3'' \times 1.4''$.  Bottom: X-ray image showing regions from
which spectra below were extracted. Blue:  North rim.
Red:  center.
\label{rx}}
\end{figure}

Figure~\ref{2xim} shows the 2009 image, smoothed with the
spatio-spectral method of \cite{krishnamurthy10}.  
Figure~\ref{rx} compares the radio and X-ray morphologies.  The marked
and surprising difference between them, particularly strong radio but weak 
X-rays in the north, suggests that the radio image
may hold a clue to the location of thermal emission.  We extracted the
spectrum from the radio-bright region shown on Figure~\ref{rx}; it is
shown in Figure~\ref{spec}.  Lines are apparent.  We found lines
as well in the central region shown in Figure~\ref{rx}, and analyzed
the regions separately.

\section{Spectral Analysis}

We combined the 2007 and 2009 datasets for a total exposure of 286 ks.
We modeled the background rather than subtracting it, and used
Markov chain Monte Carlo (MCMC) methods as implemented in the
PyMC software package \citep{patil10} to determine best parameter
values and error ranges \citep[e.g.,][]{vandyk01}.  MCMC methods
require the specification of priors on parameters to be determined (as 
described below).

\begin{figure}
\epsscale{1.1} 
%\plotone{combbrightradio.eps}
%\plotone{combbrightradioflat.jpeg}
%\plotone{combinteriormbr.eps}
%\plotone{f3.eps}
\plotone{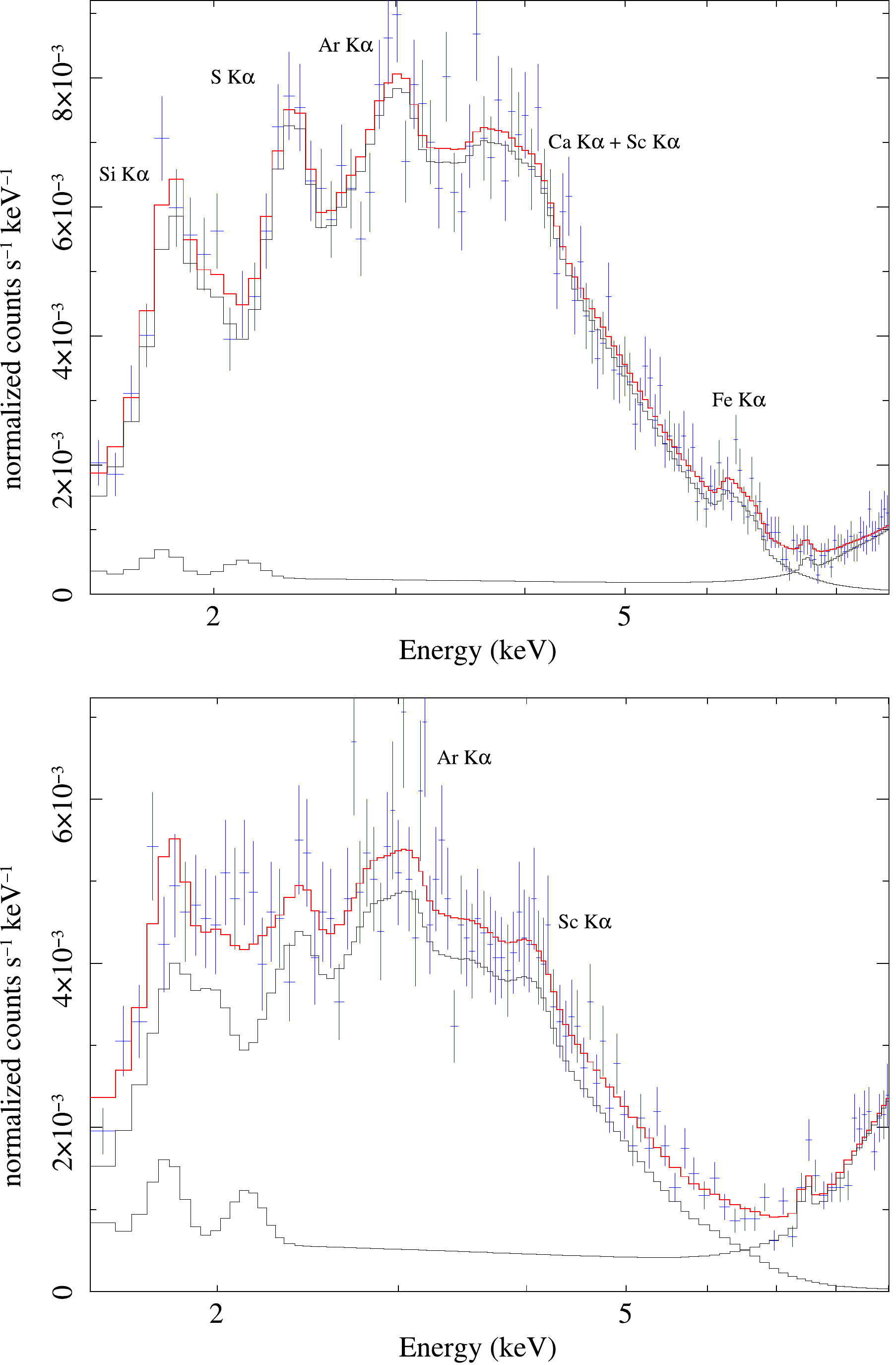}
%\plotone{combinteriormbr.jpeg}
\caption{Top: Spectrum of N rim (radio-bright region) shown in
Figure~\ref{rx}.  Lines of Ca and Sc blend together because of the
large Doppler widths.  Bottom: Spectrum of interior region.  In both
cases, background has been modeled rather than subtracted; source and
background models are described in the text.  The two black lines are
the background and source models; the red line is the total.
\label{spec}}
\end{figure}

The spectrum of the northern, radio-bright rim is shown in the upper
panel of Figure~\ref{spec}. Spectral lines typical of strongly
underionized plasma are apparent (such plasma is expected in G1.9+0.3
because of its youth and the low density of the ambient ISM). SN 1006
has a very similar X-ray spectrum; in the same spectral range of
Figure~\ref{spec}, \citet{yamaguchi08} find prominent K$\alpha$ lines
of abundant elements such as Si, S, Ar, Ca, and Fe, with line
centroids at 1.815 keV, 2.36 keV, 3.01 keV, 3.69 keV, and 6.43 keV.
(As in SN 1006, O, Ne, and Mg lines might also be present at lower energies, 
but cannot be seen because of the high absorption.) In
addition to these lines produced in hot shocked plasma, the
radioactive decay of \ti\ to \scand\ and finally to the stable isotope
$^{44}$Ca will result in the emission of X-ray and $\gamma$-ray lines
in very young remnants \citep[\ti\ decays with a mean life of $85.0 \pm
0.4$ yr,][]{ahmad06}. This decay commences via an electron capture to
\scand, leaving a K-shell vacancy followed rapidly either by Auger decay or 
by emission of a fluorescence photon of energy 4.09 keV \citep[the yield is 0.172 
photons per each \ti\ decay;][]{be04}.  Nuclear de-excitation 
Sc $\gamma$-ray lines at 78.4 and 67.9 keV are also emitted, followed by 
Ca gamma rays at 1.157 MeV (mean life of $^{44}$Sc is 5.4 hr).  An
inspection of the radio-bright rim spectrum (Figure \ref{spec})
reveals the presence of Si, S, Ar, and Fe K$\alpha$ lines, and a broad
feature near 4 keV that may be a blend of Doppler-broadened Ca and Sc
K$\alpha$ lines.  Lines are generally weaker in the low-surface
brightness interior (lower panel of Figure \ref{spec}), with Ar and Sc
lines being the most prominent.

We modeled the spectra of the northern, radio-bright rim and the faint
interior with an absorbed power law plus emission lines of Si, S, Ar,
Ca, Sc, and Fe. This simple model does not account for dust scattering
and does not separate the underlying continuum into thermal and
nonthermal components, but it suffices for the determination of line
strengths, centroids, and widths. We used a normal (Gaussian) prior
for the absorbing column density $N_H$, with mean (standard deviation) of
$6.89 (0.11) \times 10^{22}$ cm$^{-2}$, based on our multiregion
spectral fit without dust scattering to the 2007 data (Paper II). 
(Solar abundances of the absorbing ISM are those of 
Grevesse \& Sauval 1998; fits with 
dust scattering resulted in $N_H$ lower by 25\%.)  Noninformative, uniform and
logarithmic priors were assumed for the power-law index $\Gamma$ and
the (unabsorbed) 5--10 keV continuum flux $F_{5-10\ {\rm keV}}$,
respectively. 

We used normal priors for line energies, setting mean thermal line energies 
equal to the values measured by \citet{yamaguchi08} for SN 1006, and to 
4.09 keV for the Sc line.  In view of the 14,000 km s$^{-1}$ blast 
wave speed, significant bulk Doppler shifts are possible, so we chose a large 
($\sigma=10^4$ km s$^{-1}$) width for
these priors. (For numerical stability, these normal priors
were truncated to include only a finite range in line energies; we
verified that our results are not affected by this procedure.) Thermal lines 
in a young remnant arise in a fast-moving,
shocked shell bounded by forward and reverse shocks. An optically and
geometrically thin shell expanding with velocity $v_{shell}$ produces
flat-topped lines with Doppler widths of $2v_{shell}$. 
We assumed
flat-topped profiles with the same (but unknown) Doppler width for all
thermal lines. 
(Thermal broadening is likely of only modest importance, 
$\left(kT_i/m_i\right)^{1/2} \sim 0.2v_{shell}$,
based on models of 
\citet{dwarkadas98} with exponential ejecta density profiles. But it may 
still be appreciable in off-center locations such as the north rim, where 
bulk radial motions contribute less to the line broadening.)
A truncated normal prior was assumed for $v_{shell}$,
with mean of 14,000 km s$^{-1}$ and 1$\sigma$ width of 5000 km s$^{-1}$,
extending from 0 to 50,000 km s$^{-1}$. The Sc line was modeled by a
Gaussian with width $\sigma_v^{Sc}$; we assumed 
a half-normal\footnote{normal distribution with mean 0 limited to positive domain} 
prior with $\sigma = 0.15$ keV for $\sigma_v^{Sc}$. 
These priors for line widths exclude very
large ($>$ 50,000 km s$^{-1}$) widths, but otherwise provide 
weak constraints.

Some constraints on the Sc line strength are provided by IBIS/ISGRI
onboard INTEGRAL. 
The 68 and 78 keV lines have been detected by IBIS/ISGRI in Cas A
\citep{renaud06} but not in G1.9+0.3 \citep{renaud09}. We use the Cas
A detection to bound priors.  
We chose a weakly informative gamma prior for
the Sc line strength, defined as the expected line counts in
a 286 ks exposure with {\it Chandra}, with shape parameter $\alpha =
1.2$ and scale parameter $\beta = 0.01$ (see \citet{vandyk01} for
discussion of the use of gamma priors in modeling emission lines in
X-ray spectra). Its mean of $\alpha/\beta = 120$ counts is comparable
to the square root of the variance, equal to $(\alpha/\beta^2)^{1/2} =
110$ counts. This prior disfavors a strong Sc line that was searched for but 
not found in Cas A \citep{theiling06},
but
otherwise provides rather weak constraints. We assumed the same weakly
informative prior on line strength for thermal lines of Si, S, Ar, Ca,
and Fe.

The sky background was determined by a fit to the background spectrum
extracted from a large area on the S3 chip. The sky background was modeled by 
two absorbed 
power-laws, while the particle background model involved a combination
of power-laws with exponential cutoffs and narrow fluorescent lines.  
We allowed for spatial
variations in the particle background, with a logarithmic prior
imposed on the particle normalization.

Table~\ref{lineflux} contains results of spectral fits; models are
plotted in Figure~\ref{spec}. Except for fluxes, values quoted in
Table~\ref{lineflux} are standard means of the MCMC draws. We used a
geometric mean for $F_{5-10\ {\rm keV}}$, and mean line fluxes $F$
were evaluated as $(\langle F^{1/2} \rangle)^2$ \citep[according to
recommendations by][]{vandyk01}.  The 90\%\ confidence intervals were
computed using the 0.05 and 0.95 quantiles of the draws.  The Sc line
width is poorly constrained, so we provide only upper limits based on
the 90\%\ highest probability density interval for
$\sigma_v^{Sc}$.

The Sc line and thermal lines of all abundant elements have
appreciable strengths in the spectrum of the northern rim. Thermal
lines are very broad, with a full width of 28,000 km s$^{-1}$,
consistent with the estimated shock velocity of 14,000 km s$^{-1}$. The
Fe K$\alpha$ line provides the strongest constraints on line widths, but
Doppler broadening is important for other lines as well. 
Photon statistics are insufficient to study variations in Doppler widths
between different elements. This includes Sc, with a width that does not
seem to be different from thermal line widths. But Sc forms a line
blend with Ca, so errors are particularly large for these elements.
In the interior, thermal lines are much weaker; an inspection of the confidence 
intervals in Table 1 suggests that only Ar may have been 
unambiguously detected.
Notwithstanding this decline in the overall strength of
thermal lines, the Sc line remains strong (it is only a factor of 2
weaker than in the northern rim).  The prominence of Sc in the
interior is consistent with an origin in
the unshocked ejecta.

In both spectral regions independently, we examined the significance
of the Sc line detection using a likelihood ratio test as described in
\cite{protassov02}.  The null model consisted of an absorbed power law
plus Si, S, Ar, Ca and Fe lines described by flat line profiles with
equal (but unknown) Doppler widths, together with the same priors that
we used for our spectral fits.  The alternative model is obtained by
adding an additional line at 4.09 keV to the null model.  Large
samples from the posterior distribution of the null model were
obtained using MCMC simulations.  From these samples we simulated
synthetic datasets, fit each dataset with the null model and the
alternative model, and computed the likelihood ratio statistic.
The posterior predictive $p$-values for the null model
are 0.0098 and 0.030 for the northern rim and interior, respectively.
The $p$-value is equal to 0.00029 for the null model to be valid in
both of these regions.  We can therefore reject the null model, i.e.,
claim the detection of the 4.1 keV line, at 99.971\% significance.

The total Sc line flux was derived by combining the spectra shown in
Figure~\ref{spec}, and fitting the combined dataset separately with
the same spectral model as before. We obtain a total Sc line flux of
$1.2 \times 10^{-6}$ ph cm$^{-2}$ s$^{-1}$, with a 95\%\ confidence
interval of $(0.35, 2.4) \times 10^{-6}$ ph cm$^{-2}$ s$^{-1}$.  The
column density of $6.9 \times 10^{22}$ cm$^{-2}$ implies that X-rays
of 4.1 keV are attenuated by about 33\% due both to absorption and
scattering.  The Sc line fluxes have not been corrected for this, but
the correction has been made for the $^{44}$Ti masses that are
discussed next.

We expect 0.172 fluorescence photons per \scand\ atom. Thus the line 
flux, along with the source age and
distance, gives directly the amount of \ti\ synthesized in the
explosion.  Including the branching ratio for the emission of
the 1.157 MeV photon, we expect the flux of X-ray photons to
be related to that in 1.157 MeV photons by $F_x = 0.15 F_\gamma$.
The gamma-ray flux is related to the mass of \ti\ by 
\begin{displaymath}
F_\gamma = 3.6 \times 10^{-6} \frac{M({\rm Ti})}{10^{-5}\ M_\odot}
\left(\frac{D}{8.5\ {\rm kpc}}\right)^{-2} \left(\frac{\tau}{\rm yr}\right)^{-1} e^{-t/\tau} {\rm ph~cm}^{-2}  {\rm s}^{-1}
\end{displaymath}
with the mean life $\tau = 85$ yr.  (If the \ti\ is
ionized beyond the He-like state, the effective lifetime is longer \citep{mochizuki99},
but the low ionization age of Fe implied by the line centroid reported
in Table~\ref{lineflux} means that this effect should be negligible.)
We used these relations with an age of 100 yr to obtain the following
$^{44}$Ti masses: $2.8 (1.1, 5.1) \times 10^{-5} \ M_\odot$ in the
north, $1.3 (0.36, 2.6) \times 10^{-5} \ M_\odot$ in the interior, and
$3.3 (0.95, 6.5) \times 10^{-5} \ M_\odot$ for the total. (Errors are
90\% confidence limits for the two subregions, but 95\% limits for the
total.) These masses have been corrected for absorption and scattering
by being multiplied by a factor of $3/2$.  For an age of 140 yr
(corresponding to no deceleration at all),
the masses would be larger by an additional factor of 1.54.

We also modeled the spectrum of the northern, radio-bright rim with a
plane shock non-equilibrium ionization (NEI) model ({\tt vpshock} in
XSPEC).
We find $N_H = 6.88 (6.72, 7.03) \times 10^{22}$
cm$^{-2}$;
$kT_e = 3.6 \pm
0.4$ keV; and ionization timescale $\tau \equiv n_e t = 1.4 (0.78,
3.1) \times 10^{9}$ cm$^{-3}$ s.  
This extremely low ionization age is very unusual for SNRs, but
appropriate for an unprecedentedly young object.  Abundances have
substantial errors, but solar abundances are ruled out: compared to
solar, Si (set equal to S) = 3.4 (2.0, 4.6), Ar = 17 (2.8, 37), and Fe
= 4.1(2.4, 5.7).  A blueshift of line centers is required, $v_r =
-4,300 (-7,600, -1,300)$ km s$^{-1}$.  We find a FWHM of $26,500
(17,600, 34,600)$ km s$^{-1}$.
With 
$n_e \sim 2 \tau /100\ {\rm yr} = 0.7$ 
cm$^{-3}$, the estimated shocked ejecta mass 
is only $\sim 0.016 M_\odot$, with comparable masses
of intermediate-mass (Si, S, and Ar) elements and Fe.  
This is a quarter of 
the total shocked ambient gas mass of $0.06 M_\odot$ 
(for a preshock density of $\sim 0.04$ cm$^{-3}$ and a blast wave radius 
of 2.2 pc; Paper I). Impact of the ejecta with a denser than 
average ambient medium in the north may explain the relatively high shocked 
ejecta mass and density, and the strong radio emission there. 

\section{Discussion}

We summarize our results as follows:
\begin{enumerate}

\item The radio-bright N rim region shows strong K$\alpha$ lines of
Si, S, Ar, Ca, and Fe (much weaker thermal emission is also present
in the shell interior).

\item The spectral lines are both broadened and shifted.  Widths of
thermal lines are $28,000 \pm 5,000$ km s$^{-1}$ with blueshifts of
about 4,000 km s$^{-1}$.  

\item Global spectral fits with a plane-shock model imply strong
overabundances of the elements we detect, consistent with their
interpretation as ejecta dominated by intermediate-mass elements and Fe.

\item We detect for the first time with high significance a line at
  4.1 keV which we attribute to $^{44}$Sc.  The line strength of
  about $1.2 \times 10^{-6}$ ph cm$^{-2}$ s$^{-1}$ implies a mass of
  \ti\ of $3.3 (0.95, 6.5) \times 10^{-5}\, \msun$, after correction for 
  absorption and assuming an age of 100 yr and a distance of 8.5 kpc.

\item We detect \scand\ separately in the northern shell and the 
interior, with greater strength in the shell. 

\end{enumerate}

We believe the discovery of \scand\ emission to be the most important
new result.  A feature at 4.1 keV identified with \scand\ has been
reported previously in G266.2--1.2 (RX J0852.0-4622) by 
\citet{tsunemi00}, \citet{iyudin05}, and \citet{bamba05}.  However,
\citet{slane01} and, more comprehensively, \citet{hiraga09}
have failed to confirm these reports, at a level below the previous
detection claims.  It appears that G1.9+0.3 shows the first definite
detection of this transition.  Our line strength implies a flux in the
1.157 MeV line of about $1.2 \times 10^{-5}$ ph cm$^{-2}$ s$^{-1}$, well
below the COMPTEL limit of about $2 \times 10^{-5}$ ph cm$^{-2}$
s$^{-1}$ \citep{dupraz97}.  While our inferred \ti\ mass is nominally
in conflict with the upper limit of $2 \times 10^{-5}\ M_\odot$
reported by \cite{renaud09}, the errors are large enough to
accommodate it.

\ti, a trace product of explosive nucleosynthesis, carries
important information about the details of the explosion 
\citep{diehl98,the06}.
It is primarily produced in
``$\alpha$-rich freezeout'' conditions, in which rapid cooling causes
departures from nuclear statistical equilibrium and a high
concentration of free $\alpha$ particles.  In core-collapse (CC) SNe,
its production is sensitive to the location of the mass cut (within
which material becomes the neutron star), and in most models its
production is correlated with that of $^{56}$Ni, and therefore with
the supernova luminosity.  Predicted yields range from 3 to $9 \times
10^{-5} \, \msun$ \citep[see summary in][]{the06}.  Traditional SN Ia
models undergo $\alpha$-rich freezeout under somewhat different
conditions and do not produce large quantities of \ti.  However,
recent simulations of off-center delayed-detonation explosions have
shown substantially increased yields: \cite{maeda10}
quote a mass of \ti\ of $1.6 \times 10^{-5}\, \msun$, compared to $2
\times 10^{-6} \, \msun$ for a centrally-ignited pure deflagration and
$3 \times 10^{-6}\, \msun$ for a central delayed detonation.  (For
comparison, the benchmark W7 Type Ia model [Nomoto et al.~1984] gives 
$8 \times 10^{-6}\, \msun$, while a series of more
recent simulations by \citet{iwamoto99} gives a range from $(1 - 5)
\times 10^{-5}\, \msun$.)

Galactic constraints on the supernova rate can be obtained from the
observed abundance of $^{44}$Ca, and the absence of obvious discrete
sources of 1.157 MeV $\gamma$-ray emission, with the important
exception of Cas A, the only firm detection to date \citep{iyudin94,
iyudin99}.  The accompanying hard X-ray lines have also been seen from
Cas A \citep{vink01,renaud06}, with an inferred
mass of \ti\ of about $2 \times 10^{-4}\, \msun$, more than
expected from most CC models 
\citep[$(6 - 14) \times 10^{-5}\,\msun$;][]{timmes96}, though 
asymmetric models can do better
\citep[e.g.,][]{nagataki98}.

Since the 4.1 keV inner-shell transition of \scand\ formed from
electron capture in \ti\ does not require that the scandium be
ionized, we are sensitive to both shocked and unshocked material.
This is evident in the interior spectrum in Figure~\ref{spec}, in
which the \scand\ line is as prominent as in the north rim, but the
other lines are weaker.  The greater abundance of \scand\ in the
north, then, represents a true spatial distribution -- surprising,
considering that the \scand\ is expected to be produced in the
neighborhood of the Fe-peak elements, that is, the innermost ejecta.
These properties suggest that the explosion was substantially
asymmetric, perhaps consistent with the greater production of \ti\
found in asymmetric explosion models such as \cite{maeda10}.   
While He-shell denotations that trigger sub-Chandrasekhar thermonuclear 
explosions can also be invoked to explain the presence of 
asymmetrically-distributed \ti\ and Fe in the outermost ejecta layers, 
too much \ti\ is produced even in sub-Chandrasekhar explosion models 
with minimum He shell masses \citep{fink10}.
A
substantially longer observation should allow better spatial
localization of the \scand\ emission, and an unprecedented view into
supernova nucleosynthesis.

Our discovery of these spectral lines, and the \scand\ line in
particular, remains consistent with the possibility that \src\
resulted from a Type Ia event, a possibility we find increasingly
likely.  The bilaterally symmetric synchrotron--X-ray morphology, the
extremely high shock velocities we both infer and measure from line
broadening, and the prominent Fe K emission, all support a Type Ia
origin.  We have not yet ruled out a CC origin, however; while strong
iron emission is unusual in CC remnants, Cas A is a counterexample,
and it is possible that a CC origin can explain the other features as
well.  What is certainly clear is that \src\ still has a great deal to
teach us about the evolution of very young SNRs and about the
supernovae that produced them.

\acknowledgments

This work was supported by NASA through Chandra General Observer
Program grant SAO GO6-7059X.

\begin{deluxetable}{lccccccccccc}
%\rotate
\tablecolumns{12}
\tablewidth{0pc}
\tabletypesize{\footnotesize}
%\tabletypesize{\scriptsize}
\tablecaption{Spectral Fits}

\tablehead{
\colhead{Region}  &$N_H$ &$\Gamma$ &$F_{5-10\ {\rm keV}}$ &$v_{shell}$ &$\sigma_v^{Sc}$ &\multicolumn{6}{c}{Line energies (keV) and line strengths ($10^{-7}$ ph cm$^{-2}$ s$^{-1}$)} \\
\colhead{} &($10^{22}$ cm$^{-2}$) & &($10^{-13}$ ergs cm$^{-2}$ s$^{-1}$) &\multicolumn{2}{c}{($10^4$ km s$^{-1}$)} &Silicon &Sulfur &Argon &Calcium &Scandium & Iron }
%\colhead{} & & & & & &\multicolumn{6}{c}{Line strength ($10^{-7}$ ph cm$^{-2}$ s$^{-1}$)}}

\startdata
%North rim & 6.88 (6.72, 7.05) & 2.40 (2.29, 2.51) & 3.00 (2.80, 3.22) 
%&14000 (9100, 19000) & 11000 (0, 19000) &1.83 (1.78, 1.88) &2.35 (2.31, 2.40) 
%&2.99 (2.92, 3.04) &3.72 (3.59, 3.87) &6.49 (6.38, 6.61) &4.08 (4.00, 4.15) &  \\
North rim & 6.88 & 2.40 & 3.00 &1.4 & 1.1 &1.83 &2.35 
&2.99 &3.72 &4.08 &6.49  \\
& (6.72, 7.05) & (2.29, 2.51) & (2.80, 3.22) 
&(0.9, 1.9) & $<1.9$ &(1.78, 1.88) &(2.31, 2.40) 
&(2.92, 3.04) &(3.59, 3.87) &(4.00, 4.15) &(6.38, 6.61)  \\
& & & & & &4.4 &10.4 &7.7 &5.2 &10 &18  \\
& & & & & &(2.1, 7.0) &(5.9, 15.6) &(3.1, 13) &(1.2, 11) &(4.0, 19) &(11, 27)  \\

Interior & 6.89 & 2.71 & 1.35 &2.1 & 0.9 &1.83 &2.32 
&2.96 &3.74 &4.09 &6.4  \\
& (6.72, 7.06) & (2.54, 2.86) & (1.22, 1.51) 
&(1.2, 3.0) & $<1.6$ &(1.72, 1.95) &(2.17, 2.46) 
&(2.87, 3.05) &(3.52, 3.94) &(4.02, 4.17) &(6.0, 6.8)  \\
& & & & & &1.2 &3.1 &5.9 &1.9 &4.6 &3.0  \\
& & & & & &(0.1, 4.0) &(0.4, 8.3) &(1.5, 13) &(0.2, 6.0) &(1.3, 9.5) &(0.4, 8.4)  

\enddata

\tablecomments{\ Line energies (strengths) are in rows 1 and 5 (3 and 7), with 
90\% confidence limits listed in adjacent rows.}
\label{lineflux}
\end{deluxetable}

\end{document}